\documentstyle{lamuphys}
\makeatletter
\let\chapter\hid@chapter
\makeatother
\begin{document}
\pagenumbering{arabic}

\title{The TBA, the Gross-Neveu Model, and Polyacetylene}

\author{Alan\,Chodos\,\inst{1} and
Hisakazu\,Minakata\inst{2}}

\institute{Yale University, Center for Theoretical Physics,
P.O. Box 208120\\New Haven, Connecticut 06520-8120, USA
\and
Tokyo Metropolitan University, Department of Physics\\1-1  
Minami-Osawa, Hachioji, Tokyo 192-03, Japan}

\maketitle

\begin{abstract}
We summarize recent work showing how the Thermodynamic Bethe Ansatz  
may be used to study the finite-density first-order phase transition  
in the Gross-Neveu model. The application to trans-polyacetylene is  
discussed, and the significance of the results is addressed.
\end{abstract}

In this contribution, rather than simply repeat what is contained in  
the literature, we should like to attempt to place this work in  
context, and to emphasize those points that are either the most novel  
or the most inviting of further investigation.  Details of  
calculation can be found in the published references (Chodos and  
Minakata 1994, 1997).

This work brings together three disparate elements:  the Gross-Neveu  
(GN) model, the thermodynamic Bethe Ansatz (TBA), and the  
phenomenology of polyacetylene $((CH)_x)$.

Polyacetylene is essentially a linear chain which comes in two forms,  
dubbed {\it trans} and {\it cis}.  The {\it trans} form, which is the  
more stable, has a doubly-degenerate ground state.  It is this  
circumstance that allows for the existence of topological  
excitations, or solitons, and these lead in turn to a rich  
phenomenology.  An excellent review of polyacetylene and other  
similar polymers is the article in Reviews of Modern Physics by  
Heeger, Kivelson, Schrieffer and Su (1988).

It was realized long ago, by Takayama, Lin-liu and Maki (1980)  and  
by Campbell and Bishop (1982), that in the continuum limit, and in an  
approximation that ignores the dynamics of the lattice vibrations,  
the insulator-metal transition in polyacetylene can be described by  
the $N=2$ Gross-Neveu model, i.e. by the Lagrangian density

\begin{eqnarray}
{\cal L} = \bar{\psi} i \not{\delta} \psi - g \sigma \bar{\psi} \psi  
- {1 \over 2} \sigma^2  . \label{cons1}
\end{eqnarray}

This is a $1+1$ dimensional model in which $\psi$ is a 2-component  
Dirac spinor with an internal index that runs over two values (hence  
$N=2$).  It describes the electronic degrees of freedom of $(CH)_x$,  
while the auxiliary field $\sigma(x)$ represents the lattice  
distortion.  It is interesting to note, however, that the spin  
degrees of freedom of the electron are described by the internal  
index of $\psi$, while the spinor index of $\psi$ takes account of  
the right-moving and left-moving electrons near the Fermi surface. It  
is also remarkable that polyacetylene can be described, albeit  
approximately, by a relativistic model. Of course the limiting  
velocity in the Lorentz group is not the speed of light but rather  
the Fermi velocity $v_F$.

Our original motivation in this work was to understand an observed  
phase transition in polyacetylene, as a function of the concentration  
of dopants. At a critical concentration of $.06$ dopants per carbon  
atom, the data are consistent with a first-order phase transition to  
a metallic state. We attempt to describe this by computing the  
effective potential of the $GN-$model in the presence of a chemical  
potential $h$. The role of $h$ is to account for the extra density of  
electrons that is supplied to the system by the dopants. It is  
introduced by adding a term $h\psi^{\dagger}\psi$ to the Hamiltonian  
of the system.

The behavior of the $GN$ model at finite temperature and density has  
received considerable attention over the years. We note in particular  
the work of Wolff (1985), of Karsch, Kogut and Wyld (1987), of Osipov  
and Fedyanin (1987) and of Klimenko (1988). In our case, we are  
interested in the effect of the chemical potential (or electron  
density) with the temperature set to zero. We recall that, in leading  
order in the $1/N$ expansion, the zero-chemical-potential $GN$ model  
exhibits spontaneous breaking of discrete chiral symmetry (as well as  
other interesting phenomena like asymptotic freedom and dimensional  
transmutation). This corresponds nicely to the degenerate ground  
state of $trans-(CH)_x$.

As the chemical potential is increased from zero, one finds that at a  
certain value, $h = m/\sqrt{2}$, where $m$ is the  
(dynamically-generated) mass of the fermion, the system undergoes a  
first order phase transition in which the symmetry is restored.  The  
dynamical mass goes to zero above the transition point.

Remarkably enough, the critical value of $h$ that is found above  
corresponds to a dopant concentration of $.06$. Furthermore, the  
transition from the soliton-dominated broken phase to the unbroken  
phase in which the fermions behave like a free gas is consistent with  
the observed transition in which the electronic properties resemble  
those of a metal.  And the data also suggest that the transition is  
first-order (Tsukamoto 1992; Chen, et al. 1985; Moraes, et al. 1985).

The puzzling aspect of all this is why the agreement between theory  
and experiment should be so good.  Not only are we using the $GN$  
model, which is an approximation to the lattice Hamiltonian of Su,  
Schrieffer and Heeger (1980), but the results have been obtained in  
leading order in the $1/N$ expansion, and, as we have already  
remarked, the correspondence between the $GN$ model and polyacetylene  
requires $N=2$.  (In fairness, one should note that $N=2$ refers to  
$2$ Dirac fermions; in the literature one often calls this the $N=4$  
model, because there are $4$ Majorana fermions.  Thus a simple change  
of notation would appear to improve the validity of the approximation  
significantly.)

Ideally, one would like to solve the thermodynamics of the $N=2$ $GN$  
model exactly.  This is not totally out of the question, because the  
$S$-matrix, which as we shall see below is the required input for the  
$TBA$, is known exactly (Karowski and Thun 1981), and in fact work is  
in progress to solve the $TBA$ (at zero temperature) for the $N=2$  
$GN$ model numerically (Chodos, Kl\"{u}mper and Minakata, 1997).   
However, in the remainder of this note we shall discuss the problem  
of extending the analysis to the next-to-leading order in $1/N$. This  
should at least give us some indication as to whether the corrections  
might significantly affect the location, or even the existence, of  
the phase transition we have already found.

The straightforward way to go beyond leading order would be to  
compute the effective potential for the $GN$ model to next-to-leading  
order in $1/N$, including the corrections due to a non-vanishing  
chemical potential.  The analogous computation at zero chemical  
potential was performed long ago by Schonfeld (1975) and by Root  
(1975), and an inspection of their work reveals that it is already at  
a level of complexity that the labor required to incorporate $h\neq0$  
seems excessive.  We choose instead to attack the problem through the  
use of the $TBA$.

Perhaps the most interesting feature of this investigation is simply  
the discovery of how to establish the correspondence between the  
results obtained directly from the effective potential and the  
information contained in the $TBA$.  We shall find that, with the  
exception of a single dimensionless constant, we can recover all the  
information about the phase transition from the $TBA$. Furthermore,  
the unknown constant can be obtained from the effective potential  
evaluated at zero chemical potential. Once this has been understood  
and the correspondence has been verified to leading order in $1/N$,  
going to next order is simply a matter of making use of results  
already in the literature and performing one integral numerically. It  
remains an interesting question whether even the missing constant can  
be extracted from the $TBA$, so that, given the $S$-matrix the $TBA$  
would provide a self-contained and complete description of the  
thermodynamics of the $GN$ system.

The thermodynamic Bethe Ansatz, in a non-relativistic setting,  
appears in the classic paper by Yang and Yang (1969). The  
zero-temperature version  can be found in the earlier work of Lieb  
and Liniger (1963). More recent work, including the extension to  
relativistic field theory, can be found in the papers of Thacker  
(1981), Zamolodchikov (1990), Klassen and Melzer (1991), and Forgacs,  
et al. (1991a; 1991b).

At zero temperature, the essence of the $TBA$ is a linear integral  
equation for the dressed single-particle excitation energy  
$\epsilon(\theta)$.  Here $\theta$ is the rapidity of the particle:  
$E = m cosh\theta$, $p = m sinh\theta$. The equation reads:

\begin{eqnarray}
\epsilon = h - m cosh\theta + \int_{-B}^{B} d\theta^{\prime} K(\theta  
- \theta^{\prime}) \epsilon(\theta^{\prime}) \label{cons2}
\end{eqnarray}

\noindent
where the kernel $K$ is the logarithmic derivative of the $S$-matrix:

\begin{eqnarray}
K(\theta) = {1 \over 2\pi i} {d \over d\theta} lnS(\theta) .  
\label{cons3}
\end{eqnarray}

\noindent
[The $TBA$ applies to theories in $1+1$ dimension, such as the $GN$  
model, where there is only two-body scattering and the $S$-matrix is  
therefore only a phase factor depending on the relative rapidities of  
the two particles.] The parameter $B$ is determined by the condition  
$\epsilon(\pm B) = 0$ (one is implicitly assuming here first, that  
$\epsilon(-\theta) = \epsilon(\theta)$, and second, that $\epsilon$  
is positive for $-B < \theta < B$ and negative for $\mid \theta \mid  
> B$).

Once $\epsilon$ has been obtained from the solution to this equation,  
one can compute the free energy density of the system as (Forgacs, et  
al. 1991a; 1991b)

\begin{eqnarray}
f(h) - f(0) = {-m \over 2\pi} \int_{-B}^{B} d\theta \epsilon(\theta)  
cosh\theta . \label{cons4}
\end{eqnarray}

\noindent
The constant $f(0)$, on dimensional grounds, must have the form $f(0)  
= -bm^2$, where $b$ is a dimensionless constant. Our notation  
anticipates the fact that $b$ will turn out to be positive.

One notes that in the $GN$ model, the expansion of $K(\theta)$ begins  
in order $1/N$, and therefore to leading order eqn. (2) is extremely  
simple:

\begin{eqnarray}
\epsilon(\theta) = h - m cosh\theta  \label{cons5}
\end{eqnarray}

\noindent
with $B$ determined by $coshB = h/m$. We see that this is only  
possible (for real $B$) if $h \geq m$, and so we have

\begin{eqnarray}
f(h) - f(0) = {-m^2 \over 2\pi} \theta(h - m) [{1 \over 2} sinh 2B -  
B] . \label{cons6}
\end{eqnarray}

As was shown in (Chodos and Minakata 1997), one finds that the free  
energy given above can be interpreted as the value of the effective  
potential $V(\sigma)$ (properly normalized) at the point $\sigma =  
\sigma_0$ corresponding to the broken vacuum; the computation of the  
effective potential is, however, considerably more involved than the  
manipulations described above for the $TBA$, so the $TBA$ does indeed  
permit a much more efficient evaluation of the free energy.

There is, however, an essential point which demands resolution.   
Where is the evidence for a phase transition? From the effective  
potential, one learns that there is a first order transition at $h =  
m/\sqrt{2}$, but the free energy obtained from the $TBA$ is  
absolutely flat at $h = m/\sqrt{2}$, having non-trivial functional  
dependence on $h$ only for $h \geq m$.  For $h > m/\sqrt{2}$, the  
expression for the free energy obtained from the effective potential  
that agrees with the $TBA$ is not actually the value from the true  
minimum. Rather, as stated above, it is from the point representing  
the broken vacuum, which ceases to be the global minimum for $h >  
m/\sqrt{2}$.

To resolve this difficulty, one must recognize that there are really  
two phases involved: ~the massive phase characteristic of  
spontaneously broken chiral symmetry, and described by eqn. (6), and  
a massless phase corresponding to the restoration of this symmetry.   
One can obtain the free energy for this phase simply by taking the  
limit as $m \rightarrow 0$ (with $h$ fixed) of eqn. (6), recognizing  
that $f(0)$ vanishes in this limit. The result is

\begin{eqnarray}
f_0(h) = {-h^2 \over 2\pi} ~ . \label{cons7}
\end{eqnarray}

\noindent
The system will choose to be in whichever of the two phases has the  
lower free energy, and if there is a value of $h$ for which $f(h) =  
f_0(h)$, the system will undergo a phase transition at that point.

In order to make this comparison, we need to know the constant $b$  
that appears in the formula

\begin{eqnarray}
f(0) = -b  m^2 . \label{cons8}
\end{eqnarray}

\noindent
If $0 < b < 1/2\pi$, then the phase transition will occur at $h/m =  
\sqrt{2\pi b}$.

According to our present understanding, the $TBA$ itself does not  
determine $b$. However, within the effective potential formalism,  
$f(0)$ is the difference $V_0(\sigma_0) - V_0(0)$, where  
$V_0(\sigma)$ is the effective potential computed at zero chemical  
potential (up to a normalization factor of $1/N$, this is just the  
effective potential computed originally by Gross and Neveu) and  
$\sigma_0$ is the value of $V_0$ at its minimum. By consulting Gross  
and Neveu's original work (1974), one obtains

\begin{eqnarray}
b = 1/4\pi \label{cons9}
\end{eqnarray}

\noindent
and hence at the transition $h = m/\sqrt{2}$, reproducing the result  
from the effective potential.

It is now straightforward to extend this reasoning to next order in  
$1/N$. The steps are the following:

(a) One makes use of the results of Schonfeld and Root (in  
particular, in (Chodos and Minakata 1997) we employed an integral  
formula due to Schonfeld (1975)) to obtain the correction, at zero  
chemical potential, to $V_0(\sigma)$ of Gross and Neveu. One finds  
thereby that $b \rightarrow b + \Delta b$, with $\Delta b = (- {1  
\over 4 \pi}) {2.12 \over 3N}$.

(b) One inserts into the $TBA$ the first non-vanishing contribution  
to $K$, which occurs at order $1/N$. Following the work of Forgacs,  
et al. (1991a), we assume that only the $S$-matrix describing the  
scattering of fundamental fermions is relevant. [If this is not the  
case, there will be further, presumably smaller, corrections to this  
order, but we should still obtain a result that is qualitatively  
correct.]

(c) One takes the massless limit of the free energy obtained in part  
(b), giving $f_0(h) = {-1 \over 2 \pi} (1 + \delta) h^2$, with  
$\delta = ({.232 \over N})$.

One then makes the same comparison between $f(h)$ and $f_0(h)$ that  
one did in leading order, to see to what extent the phase transition  
point has been shifted.

The result for the new critical $h$ is

\begin{eqnarray}
h = {m \over \sqrt{2}} ~[1 + ({- .47 \over N})] \label{cons10}
\end{eqnarray}

\noindent
which for $N=2$ amounts to about a $20\%$ correction.  Since, in the  
massless phase, the dopant concentration is directly proportional to  
the chemical potential (see Chodos and Minakata, 1994), this implies  
that the critical concentration of dopants is about $20\%$ lower, not  
quite as good as the leading order result, but still comfortably  
compatible with experiment.

Let us conclude with a few remarks concerning a class of models for  
which the $TBA$ is exactly solvable. These models {\it a priori} have  
nothing to do with the $GN$ model that is our principal concern in  
this paper, but being exactly solvable they can provide some insight  
concerning the behavior and mathematical properties of eqn. (2).

The simplest such model has a kernel given by

\begin{eqnarray}
K(\theta) = \lambda cosh\theta  ~. \label{cons11}
\end{eqnarray}

\noindent
It is not hard to show that the corresponding $S$-matrix is $S(p_1,  
p_2) = e^{i\Phi}$, $\Phi = {2\pi \lambda \over m^2} \epsilon_{\mu\nu}  
p_1^{\mu}p_2^{\nu}$ where $
\theta = \theta_1 - \theta_2$, and $p_1$ and $p_2$ are the  
$4$-momenta of the scattered particles. This $S$ satisfies the  
appropriate restrictions imposed by analyticity, unitarity and  
crossing, but it is not polynomially bounded, and hence it is not  
clear what kind of underlying degrees of freedom are being described.

In any case, solving the $TBA$ equation one finds

\begin{eqnarray}
\epsilon(\theta) = h - \tilde{m} cosh \theta \label{cons12}
\end{eqnarray}

\noindent
where

\begin{eqnarray}
\tilde{m} = {m \over 1 + \lambda ({1 \over 2} sinh 2B - B)}  
\label{cons13}
\end{eqnarray}

\noindent
with $h = \tilde{m} cosh B$. The free energy becomes

\begin{eqnarray}
f(h) - f(0) = - \theta(h - \tilde{m}) {m^2 \over 2 \pi} ~{1 \over  
\lambda + [{1 \over 2} sinh 2B - B)]^{-1}} \label{cons14}
\end{eqnarray}

\noindent
One can see for $\lambda > 0$ that $h$ is bounded above, and in fact  
$h \rightarrow 0$ as $B \rightarrow \infty$. For $\lambda < 0$, $h$  
can take on all values, in particular $h \rightarrow \infty$ when $B  
\rightarrow B_c$ given by $1 + \lambda ({1 \over 2} sinh 2B_c - B_c)  
= 0$, but one finds that $\displaystyle{\lim_{m\rightarrow 0} f(h) =  
0}$, so there is no "asymptotically free" behavior in which the free  
energy has the form $f(h) \rightarrow -\kappa h^2$ for sufficiently  
large $h$.

If we study the next simplest model,

\begin{eqnarray}
K(\theta) = \lambda cosh 3\theta \label{cons15}
\end{eqnarray}

\noindent
[the case $K = \lambda cosh 2\theta$ is excluded by crossing  
symmetry] we again find that this is represented by a  
non-polynomially bounded $S$-matrix, and that there is a marked  
difference in behavior between positive and negative $\lambda$. The  
solution for $\epsilon$ takes the form

\begin{eqnarray}
\epsilon(\theta) = h + \epsilon_1 cosh \theta  + \epsilon_3 cosh  
3\theta  ~. \label{cons16}
\end{eqnarray}

\noindent
The coefficient $\epsilon_1$ is negative, but for $\lambda > 0$,  
$\epsilon_3$ is positive. This means that the condition $\epsilon(B)  
= 0$ is not sufficient to guarantee a solution, because there exists  
a $B^{\prime} > B$ such that $\epsilon(\theta) > 0$ for $\mid \theta  
\mid  > B^{\prime}$. We conclude that no solution exists for $\lambda  
> 0$. For $\lambda < 0$, solutions do exist that are qualitatively  
similar to the $\lambda < 0$ solutions of the previous example; i.e.,  
they do not exhibit the asymptotic freedom that would allow for the  
existence of a massless phase such as is found in the $GN$ model.

Some of the questions regarding these models are: ~(a) is it possible  
to discover what the underlying degrees of freedom are? ~(b) can one  
find an example that is more "realistic" (in the sense that it shares  
the important features of the $GN$ model)? ~(c) can one use these  
models as a laboratory for understanding how to derive the constant  
$f(0)$ directly from the $TBA$?

Finally, we conjecture that it may be possible to expand a physically  
interesting (i.e. one derived from a polynomially bounded $S$-matrix)  
kernel $K(\theta)$ in a series

\begin{eqnarray}
K(\theta) = \sum_{n=0}^{\infty}{} C_n cosh (2n + 1) \theta ~.  
\label{cons17}
\end{eqnarray}

\noindent
This would reduce the original $TBA$ equation to a matrix equation  
for the coefficients $\epsilon_n$ in the expansion

\begin{eqnarray}
\epsilon(\theta) = h + \sum_{n=0}^{\infty}{} \epsilon_n cosh (2n + 1)  
\theta  \label{cons18}
\end{eqnarray}

\noindent
which in turn might bring the problem to a more tractable  
mathematical form.

Our overall assessment of the status of this work is as follows: ~we  
have shown that the Gross-Neveu model appears to give a remarkably  
good description of the finite-density phase transition observed in  
polyacetylene, and that the result is stable against higher-order  
$1/N$ corrections. Numerical work is in progress to solve the $TBA$  
integral equation exactly for the case of interest.  Deeper questions  
remain, such as whether the $TBA$ can provide a complete account of  
the thermodynamics of the $GN$ model, and whether models can be found  
which permit the exact solution of the $TBA$ equation while at the  
same time making unambiguous physical sense.

\bigskip\noindent
{\bf Acknowledgements:} ~A.C. wishes to thank H. Meyer-Ortmanns and  
A. Kl\"{u}mper for the opportunity to attend such a stimulating  
workshop.  This work was performed under the Agreement between Yale  
University and Tokyo Metropolitan University on Exchange of Scholars  
and Collaborations (May 1996).  A.C. was supported in part by the DOE  
under grant number DE-FG02-92ER-40704.  H.M. was partially supported  
by Grant-in-Aid for Scientific Research \#09640370 of the Japanese  
Ministry of Education, Science and Culture, and by Grant-in-Aid for  
Scientific Research \#09045036 under the International Scientific  
Research Program, Inter-University Cooperative Research.

%
% ---- Bibliography ----
%

\end{document}